\documentclass[a4paper,11pt]{article}
\usepackage{jheppub}
\usepackage{amsmath}
\usepackage{amsfonts}
\usepackage{amssymb}
\usepackage{textcomp}
\usepackage{graphicx}% Include figure files
\usepackage{bm}% bold math
\usepackage{color}
\usepackage{verbatim}
\usepackage{subfig}

\usepackage{epsfig}
\usepackage{amsfonts,amsmath,amssymb}

\newcommand{\be}{\begin{equation}}
\newcommand{\ee}{\end{equation}}
\newcommand{\ba}{\begin{eqnarray}}
\newcommand{\ea}{\end{eqnarray}}
\newcommand{\ben}{\begin{enumerate}}
\newcommand{\een}{\end{enumerate}}

\newcommand{\p}{\partial}

\newcommand{\la}{\langle}
\newcommand{\ra}{\rangle}

\newcommand{\rar}{\rightarrow}

\begin{document}

\preprint{VNIIA-CFAR- 01/12}

\title{Holographic model of the $S^\pm$ multiband superconductor}

\author{A. Krikun}
\author{V. P. Kirilin}
\author{A. V. Sadofyev}

\affiliation{All-Russia research institute of automatics (VNIIA), \\
Center for fundamental and applied research (CFAR),  \\
 Moscow, Russia \\
%22, ul. Sushchevskaya, Moscow 127055, Russia
}

\affiliation{Institute for Theoretical and Experimental Physics (ITEP), \\ Moscow, Russia 
%B. Cheryomushkinskaya 25, 117218 Moscow, Russia
}

\abstract{
We construct the holographic model of an $S^\pm$ multiband superconductor. This system is a candidate to explain the anomalous features of the iron-based superconductors (e.g. LaFeAsO, BFe2As2, and other pnictides and arsenides). We study the framework, which allows formation of the sign-interchanging order parameter. We also calculate the electric AC conductivity and study its features, related to the interband interaction.  
}

\maketitle

\section{Introduction}
After its recent discovery the family of superconductors based on the iron compounds ($LaFeAsO$, $BaFe_2As_2$ etc.) has attracted lots of attention and revived interest to unconventional superconductivity. The question about a possible mechanism of superconductivity is intimately related to the symmetry of the order parameter in the system and there is still no consensus about the form of this symmetry in iron superconductors \cite{Fe_review1, Fe_review2}. These compounds all share the characteristic multiband feature. Namely they possess multiple conductive bands, which cross the Fermi energy level along several Fermi surfaces. In the DFT calculations and ARPES experiments it was shown that iron-based superconductors have at least a pair of electron and hole Fermi surfaces in the middle ($\Gamma$ point) and in the corner (M point) of the Brillouin zone, respectively\footnote{There are, however, exclusions, such as $KFeSe$, which has only electron-type Fermi surfaces. \cite{KFeSe}}. Such a band structure allows formation of the different superconducting condensates on different conductive bands, what leads to a wide variety of possible symmetries of the superconducting order parameter. Among them one of the simplest possibilities was considered in \cite{Mazin1, Mazin2}. If the material has two different conductive bands (populated with holes and electrons, respectively), it is possible that two charged isotropic (S-wave) condensates are formed on them: $\Delta_1$ and $\Delta_2$. There are possibilities where they have the same ($sign(\Delta_1) = sign(\Delta_2)$) or opposite ($sign(\Delta_1) = -sign(\Delta_2)$) signs and these arrangements are called $S^{++}$ and $S^{\pm}$ order parameters, respectively. The two-band system can be studied within the Bardeen-Cooper-Schrieffer, Eliashberg or Ginzburg-Landau approaches \cite{Mazin1, Mazin2, Mazin_review, Golubov, Kagan} and one can show that the $S^{\pm}$ superconductivity has a number of interesting properties. Namely it can be realized even if the intraband interaction is weak and the interband interaction is repulsive. In the case of different signs of the order parameters at the different bands the interband interaction may be mediated by the spin fluctuations, and thus the mechanism of the Cooper pair creation may be magnetic rather then phononic. Moreover, one should anticipate the rich spin dynamics in iron-based superconductors, as they exhibit antiferromagnetic instability just near the superconducting region of the phase diagram. A strong interband interaction in these systems is also usually related to the anomalous features in the AC (optical) electrical conductivity -- the peaks in the mid-infrared region \cite{Peaks,exp_BaFeAs,exp_LaFeAso}.

Since the underlying dynamics of superconducting system under consideration is presumably strongly coupled it is reasonable to try to construct a holographic model of multiband superconductor with an $S^{\pm}$ symmetry of the order parameter, which might describe the variety of experimental data and provide some insight into the relations between different features of unconventional superconductors. One of the first holographic models of superconductivity has been proposed in \cite{hartnoll1, hartnoll2, Gubser1}. It described the one-band $S$-wave superconductor and has shown the effectiveness of the method by correctly reproducing results of the Ginzburg-Landau theory and predicting the nontrivial dependence of AC-conductivity on the frequency. Later other models have been constructed for P-wave \cite{Gubser2, Gubser3} and D-wave \cite{d-wave0,d-wave1,d-wave2} superconductors, describing the vortex and droplet formation \cite{vortex, vortex2} and taking into account the effects of crystal lattice \cite{Tong1, lattice2, Tong2} and impurities \cite{impurities} \footnote{Reviews of the subject include \cite{hartnoll_lectures, mcgreevy_lectures, herzog_lectures, sachdev_review, benini_rewiew}}. In our present study we will also use the approach similar to \cite{multiband} where the multiband superconductor was considered, although our construction is different from that used in \cite{multiband} providing us with the possibility to treat the phases of condensates and electric current straightforwardly. The paper is organized as follows: in Section II we introduce the holographic model of $S^{\pm}$ superconductor, in Section III we study the phase diagram of the model and the features of different condensate patterns, Section IV is devoted to the calculation of the electric conductivity of the dual superconductor, which possesses anomalous features related to the interband interaction, the conclusion is given in Section IV. Appendix A is devoted to the derivation of the full set of equations of motion in our model. In Appendix B we elaborate the numerical equation of state.

\section{Holographic model}

We start building the holographic model of $S^{\pm}$ superconductor by identifying the degrees of freedom in the bulk theory. First of all, the $S$ symmetry of the order parameter suggests that its dual bulk field should be a scalar \cite{hartnoll1,hartnoll2}. Moreover, in the two-band material that we are considering there are two possible gaps $\Delta_i (i=1,2)$ in different bands, as we pointed out earlier, each with its own complex phase. This allows us to argue that there may be a hidden global symmetry between bands in the boundary theory. The use of such symmetry is the main ingredient of our model, which leads to the results qualitatively similar to experimentally observed phenomena. Thereby we organize the two gaps in the fundamental representation of a $U(2)$ group. This global symmetry in the boundary theory leads, according to the general holographic principle, to the similar gauge symmetry in the gravity side of the correspondence. Thus, in the bulk we obtain the fundamental scalar field $\phi = (\phi_1 , \phi_2)^T$ charged under the nonabelian $U(2)$ gauge symmetry. The method of describing multiband holographic systems by use of the hidden global symmetry was adopted also in \cite{multiband}, but the usage of $U(2)$ group instead of $SO(3)$ allows us to define phases of complex condensates separately, what is otherwise rather tricky. The action of the bulk theory is
\begin{gather}
\label{action}
S = \int d^3x dr \ \sqrt{-g} \left[- \frac{1}{4} tr( F_{\mu \nu} F^{\mu \nu}) - (D_\mu \phi)^\dag (D_\mu \phi) + 2 \phi^\dag \phi \right],
\end{gather}
where $F_{\mu \nu} = \p_\mu A_\nu - \p_\nu A_\mu - i [A_\mu, A_\nu]$, $A_\mu$ is the $U(2)$ gauge field, the covariant derivative is $D_\mu \phi^i = \p_\mu \phi^i + i (A_\mu \phi)^i$ and we rescaled the bulk gauge coupling to unity. We choose the mass of the scalar field according to the single band case \cite{hartnoll1}: $m^2_\phi = -\frac{2}{L^2}$. We will use the probe approximation, neglecting the backreaction of the bulk fields on the gravity, so the background metric is just the AdS-Schwarzschild black hole:
\begin{equation}
\label{metric}
ds^2 =  -f(r) dt^2 + \frac{dr^2}{f(r)} + r^2(dx^2 + dy^2).
\end{equation}
The blackening factor $f(r) =\frac{r^2}{L^2} - \frac{r_h^3}{r L^2}$ defines the position of the black hole horizon $r_h$, which is related to the temperature of the boundary theory as
\begin{equation}
\label{temp}
T = \frac{3 r_h}{4 \pi L^2}. 
\end{equation}
Throughout the paper we will always rescale the length units so that $L = 1$. 

In what follows we try to find the solution to the equations of motion derived from (\ref{action}), which describes the condensation of the charge carriers in both bands of the boundary theory. In accordance to the holographic dictionary the condensate $\Delta$ of the operator is proportional to the normalizable mode in the $r \rar \infty$ expansion of the classical solution for the corresponding bulk field $\phi$. 
\begin{equation}
\phi(r)\Big|_{r \rar \infty} = \frac{\phi^{(1)}}{r} + \frac{\phi^{(2)}}{r^2} + \dots
\end{equation}
In the present case both modes are normalizable and we choose the second one $2 \Delta = \sqrt{\sqrt{2} \phi^{(2)}}$ (the normalization follows \cite{hartnoll1}) thus defining a particular boundary system to be described (see \cite{hartnoll1, hartnoll2} for discussion of possible choices). The first mode $\phi^{(1)}$ is related to the source of the operator in the boundary theory and should be put to zero as there are no such sources in the problem under consideration. To find the solution with two condensates we use the ansatz:
\begin{equation}
\label{ansatz}
\phi(r) = \phi_0(r)\begin{pmatrix} \cos(\theta) \\ \sin(\theta)  \end{pmatrix}, \qquad A = M(r) \begin{pmatrix} 1 & 0 \\ 0 & 1 \end{pmatrix} dt  + \Lambda(r) \begin{pmatrix} 0 & 1 \\ 1 & 0 \end{pmatrix} dt,
\end{equation}
with constant $\theta$, which describes various possible relations between two condensate. We find (see Appendix A) that to satisfy the equations of motion of the model in given ansatz $\theta$ must assume the values
\begin{align}
\label{theta-values}
\theta^+ =& (4n+1) \frac{\pi}{4}, (n \in \mathbb{Z}), \quad \qquad \mbox{$S^{++}$ state,} \\
\notag
\theta^- =& (4n+3) \frac{\pi}{4}, (n \in \mathbb{Z}), \quad \qquad \mbox{$S^{\pm}$ state.}
\end{align}
Actually the state, in which the scalar field condenses, depends strongly on the nonabelian component of the gauge field turned on in the particular ansatz, as it is proportional to corresponding eigenvector. We will comment on the other possible solutions bellow. Here the $\theta$ phases with superindex $+$ and $-$ describe the $S^{++}$ ($\Delta, \Delta$) and $S^{\pm}$ ($\Delta, -\Delta$) symmetries of the order parameter, respectively. The time component of the Abelian gauge field $M(r)$ is dual to the number of particles operator in the boundary theory $c_i^\dag c_i$, hence its boundary value is equal to the corresponding chemical potential. In this ansatz we consider the equal chemical potentials on two bands ($\mu_1 = \mu_2 = \mu$), assuming that the electron and hole bands of the superconductor are doped by the same amount of carriers. It is straightforward to generalize this model to the case of different chemical potentials in the bands by introducing the third component of the nonabelian gauge field ($A^3 \sim diag(1,-1)$), but we won't do this here for simplicity. The off-diagonal component of the nonabelian gauge field $\Lambda$ is also dual to the scalar current in the boundary theory similar to the particle number operator, but this time it describes the jumps of particles from one band to another: $c^\dag_1 c_2 + c^\dag_2 c_1$. The presence of the source for such an operator would mean that there exists an external potential, which results in the mixture of different bands. Thus only mixed states, which diagonalize the matrix of ``chemical potentials'', have definite particle numbers. 

We should note here that our model possesses the nonabelian gauge field, which is the same as described in \cite{Gubser3} in the context of the $P$-wave superconductivity. In \cite{Gubser3} it was shown that even without scalars the spatial component of the gauge field tends to condense at certain temperature. We stress, however, that for the $P$-wave condensation described in \cite{Gubser3} the nonabelian charge of the black hole is needed (It is related to the corresponding chemical potential). Contrary, in our model the chemical potential is introduced for the Abelian subgroup of U(2) and black hole possesses only the Abelian charge. Thus the nonabelian gauge field can be sourced by this charge only via the interaction term involving the scalar. Consequently, in case of the purely Abelian chemical potential the situation described in \cite{Gubser3} is not realized and the only possible phase transition is that involving the scalar condensation. This situation does not change qualitatively even if we introduce the small nonabelian chemical potential $\lambda$, as described below. The $P$-wave condensation becomes possible, but the critical temperature of this phase transition is defined by the value of the $\lambda$, and as long as we consider the nonabelian chemical potential much smaller than the Abelian one ($\lambda \ll \mu$) the scalar condensation occurs at higher temperatures. It would be, nevertheless, very interesting to study the system with comparable chemical potentials as it should demonstrate the competing $S^\pm$ and $P$-wave orders, but this subject is beyond the scope of the present paper.

In Appendix A we check that the ansatz (\ref{ansatz}) satisfies the full set of equations of motion following from (\ref{action}) if the functions $\phi_0(r), M(r), \Lambda(r)$ are solutions to (primes denote derivatives with respect to $r$)
 \begin{align}
\label{EOMS}
\phi_0:& \qquad  \phi_0'' + \left( \frac{f'}{f} + \frac{2}{r} \right) \phi_0' + \frac{(M \pm \Lambda)^2}{f^2} \phi_0 + \frac{2}{f(r)} \phi_0 = 0, 
\\
\notag
M:& \qquad M'' + \frac{2}{r} M' - \frac{\phi_0^2}{f} (M \pm \Lambda) = 0, 
\\
\notag
\Lambda:& \qquad  \Lambda'' + \frac{2}{r} \Lambda'   \mp \frac{\phi_0^2}{f} (M \pm \Lambda) =0. 
\end{align}
Here and below upper signs correspond to $S^{++}$ ansatz and lower to $S^{\pm}$. One can combine the equations for $M$ and $\Lambda$ in such a way that the system will look exactly like the equations of motion for a single scalar studied in \cite{hartnoll1}
\begin{align}
\label{EOMS2}
\phi_0:& \qquad  \phi_0'' + \left( \frac{f'}{f} + \frac{2}{r} \right) \phi_0' + \frac{(M \pm \Lambda)^2}{f^2} \phi_0 + \frac{2}{f(r)} \phi_0 = 0, 
\\
\notag
M \pm \Lambda:& \qquad (M \pm \Lambda)'' + \frac{2}{r} (M \pm \Lambda)' - \frac{2 \phi_0^2}{f} (M \pm \Lambda) = 0. 
\end{align}
This system describes a scalar field that gets a nonvanishing normalizable component at low temperatures $T < T_c$. The other linear combination of the equations (\ref{EOMS}) is
\begin{align}
\label{EOM3}
M \mp \Lambda:& \qquad (M \mp \Lambda)'' + \frac{2}{r} (M \mp \Lambda)' = 0. 
\end{align}

In \cite{hartnoll1} the critical temperature $T_c$ was related to the density of charge carriers, the normalizable component of the field $(M(r) \pm \Lambda(r))$. In our case the charge carriers density $\rho$ and normalizable component of $(M(r) \pm \Lambda(r))$ are generally not equal, so we should study the equation of state for our system in more detail. First of all let us fix the notation, denoting the asymptotic coefficients for the fields as
\begin{align}
\label{asympt}
\Lambda(r) \Big|_{r \rar \infty} &= \lambda - \frac{J}{r}, \\
\notag
M(r) \Big|_{r \rar \infty} &= \mu - \frac{\rho}{r}.
\end{align}
Here $\rho$ is proportional to the charge carriers density $\la c_i^\dag c_i \ra$, as by holographic prescription the normalizable mode of the field corresponds to the mean value of dual operator. Similarly, $J$ is proportional to the value $\la c_1^\dag c_2 + c_2^\dag c_1 \ra$, which can be interpreted as the particle jumping rate between the bands. 

\begin{figure}[h!]
\centering
\includegraphics[width=0.5 \linewidth]{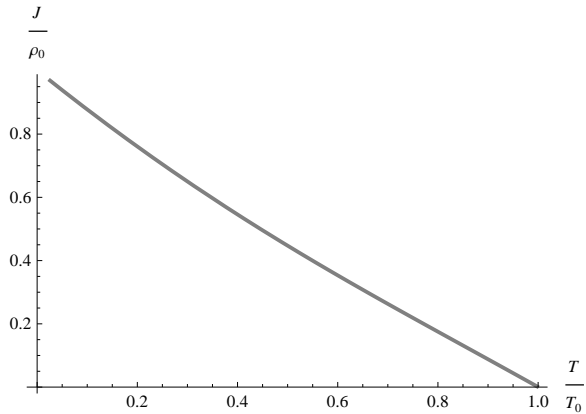}
\caption{\label{current} The current $J = \la c_1^\dag c_2 + c_2^\dag c_1 \ra$ in $S^{++}$ condensed state of two-band superconductor (in $S^{\pm}$ it has opposite sign) in absence of the corresponding source $\lambda$.}
\end{figure}

As the system of equations (\ref{EOMS2}),(\ref{EOM3}) is scale invariant, we can choose one of the parameters of our model to define the scale of all dimensional quantities. Taking $\mu = \mu_0 \sim const$ would mean we are considering grand canonical ensemble and the number of particles can change with temperature. The second possibility is to take  $\rho = \rho_0 \sim const$ as reference scale, considering the canonical ensemble with a constant number of particles on the bands. In what follows, we assume the latter point of view. Moreover, we find it convenient to keep constant the ratio of $\mu$ and $\lambda$ in order to control the mixing angle of band states.
\begin{equation}
\alpha = \frac{\lambda}{\mu} \sim const
\end{equation}

Equations (\ref{EOMS2}),(\ref{EOM3}) unambiguously define the state of the system. The procedure of getting the corresponding parameters from the numerical solution is described in Appendix B. In the end of the day we find that the solution with nonzero scalar condensate exists in both ans\"{a}tze (\ref{ansatz}) below certain critical temperature, which is (see \ref{critical_T})        
\begin{align}
\label{T_c}
T_{c} & \approx 0.118 \sqrt{ \rho_0 (1+\alpha)} \quad \mbox{for $S^{++}$ state,} \\
\notag
T_{c} & \approx 0.118 \sqrt{ \rho_0 (1-\alpha)} \quad \mbox{for $S^{\pm}$ state.}
\end{align}

It is also noteworthy that in the condensed phase of our two-band superconductor there is a nonzero current $J = \la c_1^\dag c_2 + c_2^\dag c_1 \ra$ even in the absence of external source $\lambda=0, \alpha=0$ (see Fig. \ref{current}). This can be interpreted as the mediation of the strong interband interaction in the bulk, which  leads to the interband jumping of the charge carriers.

\section{Phase diagram} 
One can see that in the situation when the off-diagonal component of the chemical potential $\lambda$ is absent the two possible configurations of the condensates are degenerate: they have the same critical temperature (\ref{T_c})  and the same free energy, as the equations (\ref{EOMS2}) for different ans\"{a}tze switch places when the sign of $\Lambda$ is changed. Consequently, any perturbation of the system would break this degeneracy and we can consider nonzero $\lambda$ as one of the possibilities. One should note, however, that if  $\lambda = 0$ then there is no reason the ansatz (\ref{ansatz}) would have minimal free energy among other possible solutions. Indeed, in such a situation the solution with nontrivial bulk profile of any nonabelian gauge field component should have the same free energy as (\ref{ansatz}) by the symmetry reasons. But as we have already noted, the condensation pattern of the scalar field (and consequently the symmetry of the superconducting gap) is crucially dependent on the type of condensed gauge component, as it is proportional to one of its eigenvectors. Namely, in order to exhibit the gap in both conductive bands the system should contain a perturbation, which leads to the condensation of $A^1$ component of the gauge field. Hence, turning on the potential $\lambda$ is a reasonable choice to achieve this effect. In this section we will check the free energy of different condensation patterns at nonzero $\lambda$ to find out, which solution is the stable one.

\begin{figure}[h!]
\centering
	\begin{minipage}[h]{0.45\linewidth}
	\includegraphics[width=0.9 \linewidth]{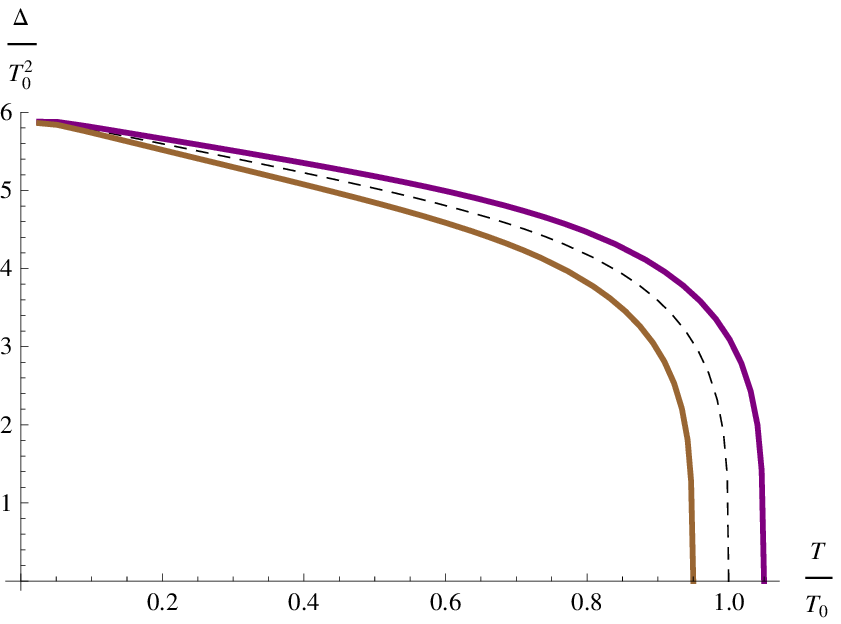}
	\caption{\label{order}Dependence of the gap value on the temperature for $S^{\pm}$ state (purple curve) and $S^{++}$ state (brown curve) at negative $\lambda=-0.1 \mu$. Dashed curve -- degenerate state at $\lambda=0$.}
	\end{minipage}
\ \ \           
	\begin{minipage}[h]{0.45\linewidth}
	\includegraphics[width=0.9 \linewidth]{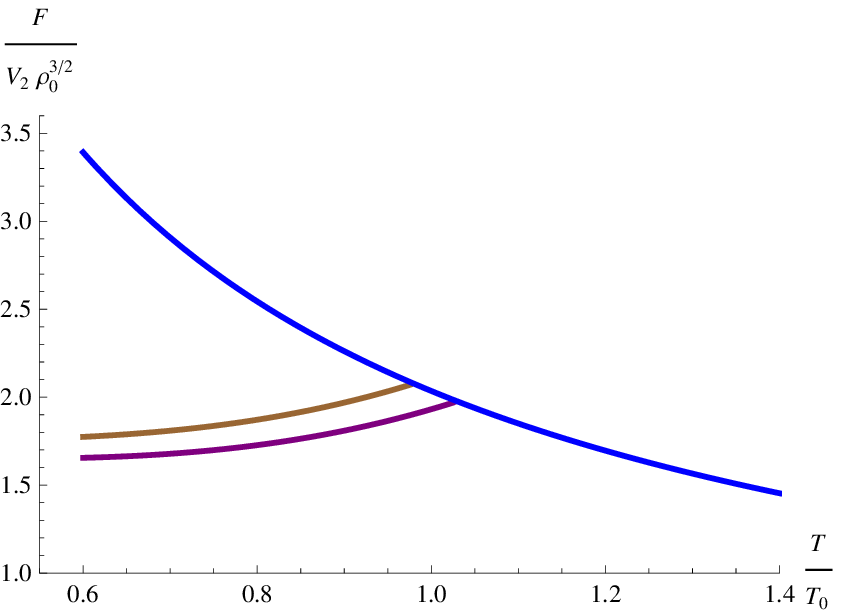}
	\caption{\label{free_energy}Free energy of $S^{\pm}$ state (purple curve) and $S^{++}$ state (brown curve) at $\lambda=-0.05 \mu$. Blue curve -- free energy of the normal (nonsuperconducting) phase at the same $\lambda$.}
	\end{minipage}
\end{figure}

First of all we note, that the critical temperatures of $S^{++}$ and $S^{\pm}$ differ at $\alpha \neq 0$. As we would like to study the $S^{\pm}$ condensation, in what follows we will take $\alpha = -0.05$, corresponding to small negative $\lambda$. Hence on the plot of the order parameter value with respect to the temperature (normalized to the critical temperature $T_0$ at $\lambda=0$) Fig.\ref{order}(a) we see that the critical temperature of $S^{\pm}$ phase in this situation is higher. 

To find out what configuration of the order parameter is stable we compute the free energy $F$ of our solutions. By the holographic principle $F$ is related to the on-shell action for given solution. The action on our ansatz (\ref{ansatz}) is
\begin{eqnarray}
  S_{on-shell}=\int d^3x dr r^2\left[ (\partial_r M) ^2+\left(\partial_r\Lambda\right)^2-f\left(\partial_r\phi_0\right)^2+\frac{1}{f} \phi_0^2 \left(M\pm\Lambda\right)^2+2\phi_0^2\right].
\end{eqnarray}
Integrating by parts and using the equations of motion we can rewrite it in the form
\begin{eqnarray}
  S_{on-shell}=V_2 \int dt \left[  r^2 \left(M\partial_r M+ \Lambda \partial_r \Lambda- f \phi_0 \partial_r \phi_0 \right)\Big|^\infty_{r_h}- \int \phi_0^2\left(M\pm\Lambda\right)^2\frac{r^2}{f(r)} dr \right],
\end{eqnarray}
where $V_2$ is 2d volume. This yield for the $F = TS_{on-shell}$ :
\begin{eqnarray}
F=V_2 \left[ (\mu\rho+\lambda J) - \int \phi_0^2\left(M\pm\Lambda\right)^2\frac{r^2}{f(r)}dr \right].
\end{eqnarray}
We can clearly interpret the first term as the value of $F$ in the normal, uncondensed phase. The second one is always negative, rendering the superconducting phase thermodynamically favorable. We observe also that the condensate with higher $T_c$ for given $\lambda$ is the one, which persists when we further lower the temperature, since it has lower $F$. That is, for $\lambda > 0$, $T_{c}^{++} > T_{c}^{\pm}$ and $F^{++} < F^{\pm}$, and vice versa for $\lambda < 0$. Indeed we see the corresponding behavior on the numerical plot of $F$ for the normal and condensed phases on Fig.\ref{free_energy}.

\begin{comment}

\section{Solution without nonabelian fields}
 
Another interesting ansatz to be considered is the ansatz without any interaction between bands, namely
\begin{equation}
\label{ansatz2}
\phi = \begin{pmatrix} \phi_1 (r)  \\ \phi_2(r) \end{pmatrix}, \qquad A = \mu(r) \begin{pmatrix} 1 & 0 \\ 0 & 1 \end{pmatrix} dt 
\end{equation}
In this ansatz the equations of motion reduce to
\begin{align*}
\phi_0: \qquad  &\frac{1}{\sqrt{-g}} \p_\mu \sqrt{-g}  g^{\mu \nu} \p_\nu \phi_0 - \phi_0 \Big[ (\hat{A}_\mu \big)^2  -2 \Big] = 0 
\\
\hat A_\mu: \qquad &\frac{1}{\sqrt{-g}} \p_\nu \sqrt{-g}  g^{\mu \rho} g^{\nu \lambda}  \hat{F}_{\lambda \rho} -  \phi_0^2 g^{\mu \rho} \Big( \hat{A}_\rho + \p_\rho \alpha \Big)  = 0 
\\
A^1_\mu:    \qquad &  \phi_0^2 \sin(2 \theta) \cos(2 \gamma)  g^{\mu \rho} \big(\hat{A}_\rho + \p_\rho \alpha \big) =0 
\\
A^2_\mu:    \qquad &  \phi_0^2 \sin(2 \theta) \sin(2 \gamma)  g^{\mu \rho} \big(\hat{A}_\rho + \p_\rho \alpha \big)=0 
\\
A^3_\mu:    \qquad & \phi_0^2 \cos (2 \theta) g^{\mu \rho} \big(\hat{A}_\rho + \p_\rho \alpha \big)   =0 
\end{align*}
Suddenly, this solution does not exist!!!

\end{comment}

\section{Electrical conductivity}
First of all in order to compute the electrical conductivity in the condensed phase of $S^{\pm}$ holographic superconductor we identify the electromagnetic current in the model. As the two conductive bands are populated with the carriers with opposite charges (electrons and holes), the electromagnetic field will couple differently with each of the condensates. Namely, the electromagnetic current in this model is dual to the nonabelian component of the gauge field
\begin{equation}
 J_\mu^{e.m.} \leftrightarrow  A^3 \begin{pmatrix} 1 & 0 \\ 0 & -1 \end{pmatrix},  \qquad  A^3(r) \Big|_{r \rar \infty} = A^{3(0)} + \frac{A^{3(1)}}{r}. 
\end{equation}
Thus, in order to study the electric conductivity we consider the perturbation of $A^3$ on the background of the condensed solution and calculate the relation between its normalizable and non-normalizable components. The former is related to the electric current in the boundary theory and the latter to the vector potential of external electric field:
\begin{equation}
\sigma_{ij}(\omega) = \frac{\la J^{e.m.}_i \ra}{i \omega A^{e.m.}_j} = \frac{1}{i \omega} \frac{A_i^{3(1)}}{A_j^{3(0)}}.
\end{equation}
Let us turn on the electric field along the $x$-axis. By looking at the equations of motion (see Appendix A) one can notice that $A_x^3$ multiplied by $A_t^1 = \Lambda(r)$ provides a linear source for the other gauge field component $A_x^2$. In its turn $A_x^2$ enters the equation of motion for $A_x^3$. Thus, about the background (\ref{ansatz}) we get the system of coupled linear equations on $A_x^3, A_x^2 \sim e^{i \omega t}$, which describes the perturbation of the electric field
\begin{align}
A^3_x:    \qquad  \p_r^2 A_x^3 + \frac{f'}{f} \p_r A_x^3 +  \left(\frac{\omega^2}{f^2}  - \frac{\phi_0^2}{f} + 4 \frac{\Lambda^2}{f^2}  \right) A_x^3 + \frac{4 i \omega \Lambda}{f^2} A_x^2   = 0,  \\
A^2_x:    \qquad  \p_r^2 A_x^2 + \frac{f'}{f} \p_r A_x^2 +  \left(\frac{\omega^2}{f^2}  - \frac{\phi_0^2}{f} + 4 \frac{\Lambda^2}{f^2}  \right) A_x^2 - \frac{4 i \omega \Lambda}{f^2} A_x^3   = 0.  
\end{align}
It can be readily rewritten as a couple of equations on the complex functions $\mathcal{A}^\pm = A_x^3 \pm i A_x^2$ 
\begin{align}
\p_r^2 \mathcal{A}^\pm + \frac{f'(r)}{f(r)} \p_r \mathcal{A}^\pm +  \left[\frac{\big(\omega \pm 2 \Lambda(r) \big)^2}{f(r)^2}  - \frac{\phi_0^2}{f(r)} \right] \mathcal{A}^\pm = 0.
\end{align}
Now in order to compute the conductivity we solve these equations with the following boundary conditions. On the horizon they correspond to the wave, which falls into the black hole: $\mathcal{A}^{+},\mathcal{A}^{-}  \sim (r-r_h)^{\frac{i \omega}{3 r_h}}$. And on the AdS boundary ($r\rar \infty$) they describe the equal sources: $\mathcal{A}^{+}=\mathcal{A}^{-} = A_x^{3(0)}$. Then we compute the relation of the normalizable and non-normalizable modes of $A_x^3(r) = \frac{1}{2} \big(\mathcal{A}^{+}(r) + \mathcal{A}^{-}(r)\big)$. 

\begin{figure}[h!]
\centering
  \subfloat[][]{
	\begin{minipage}[h]{0.5\linewidth}
	\includegraphics[width=0.9 \linewidth]{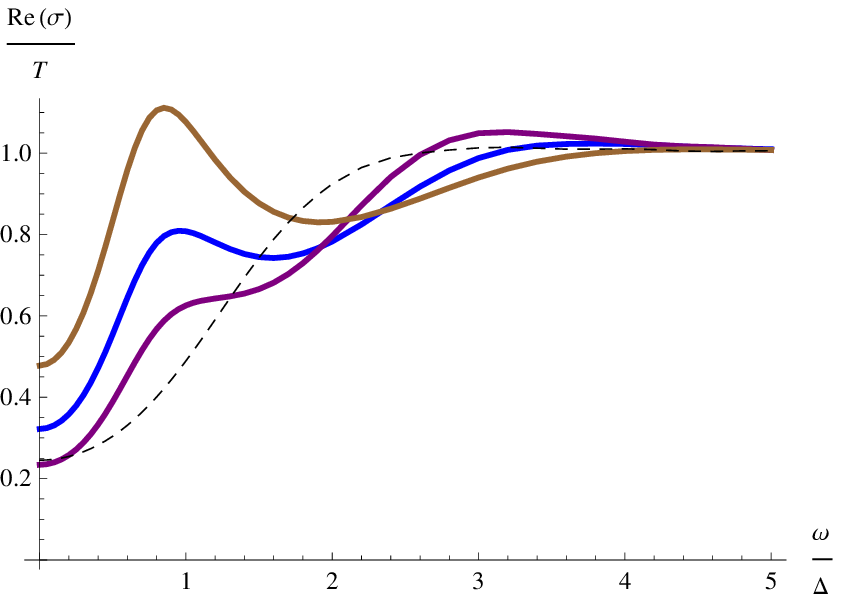}
	\end{minipage}}                
  \subfloat[][]{
	\begin{minipage}[h]{0.5\linewidth}
	\includegraphics[width=0.9 \linewidth]{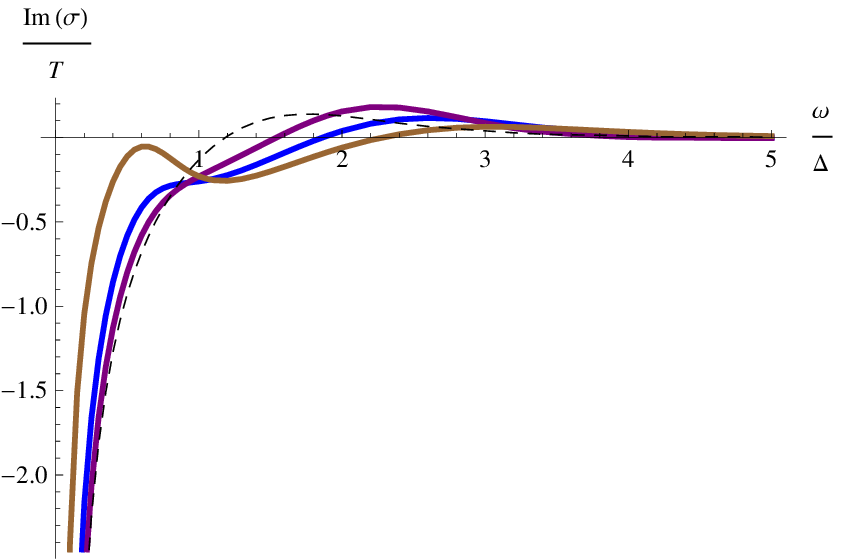}
	\end{minipage}} 
\caption{\label{conduct} The real (a) and imaginary (b) parts of the AC conductivity depending on frequency at negative $\lambda= -0.05 \mu$:  purple curve -- $S^{\pm}$ state, brown curve -- $S^{++}$(this state is quasistable at given $\lambda$). Blue curve -- degenerate case with $\lambda=0$. Dashed black curve -- the ordinary one-band superconductor with the same parameters. Temperature is constant $T \approx 0.8 T_0$ }
\end{figure}

The dependence of the AC conductivity on the frequency for constant temperature \hbox{($T \approx 0.8 T_0$)} for the $S\pm$ and $S^{++}$ ans\"{a}tze is plotted on Fig.\ref{conduct}. The imaginary part of conductivity exhibits a pole at $\omega =0$ in all cases, what is a signal of superconductivity  and is related to the $\delta$-function at $\omega =0$ in the real part of $\sigma$ by the Kramers-Kronig relation. This pole ensures us that our model describes indeed a superconductor.
More interestingly, even without introducing the external interband potential $\lambda=0$ (blue curve) we see qualitative change of the behavior of the real part of the conductivity in comparison with the ordinary one-band superconductor with the same parameters (dashed curve).  When we consider finite off-diagonal chemical potential $\lambda = -0.05 \mu$ the peak in the conductivity at $\omega \sim \Delta$ lowers in $S^{\pm}$  superconductor (purple curve) and rises in $S^{++}$ superconductor (brown curve). One should note however that $S^{++}$ state is quasistable at such $\lambda$ (see previous Section), so in the real material the state with lower peak will dominate. Based on its sensitivity to the $\lambda$ we can conclude that this feature is tightly related to the interband interaction, so its appearance should be observed in strongly coupled multiband systems. From the other hand, the sufficient external potential for interband current, which can be caused for instance by appropriate impurities, leads to the damping of the peak, so that its only effect is in changing the slope of conductivity in the mid-infrared region to almost linear. This situation can take place in the iron-based superconductors, where no broad mid-infrared peak was observed, but the slope of conductivity is anomalous \cite{charnukha}.

\begin{figure}[h!]
\centering
  \subfloat[][]{
	\begin{minipage}[h]{0.5\linewidth}
	\includegraphics[width=0.9 \linewidth]{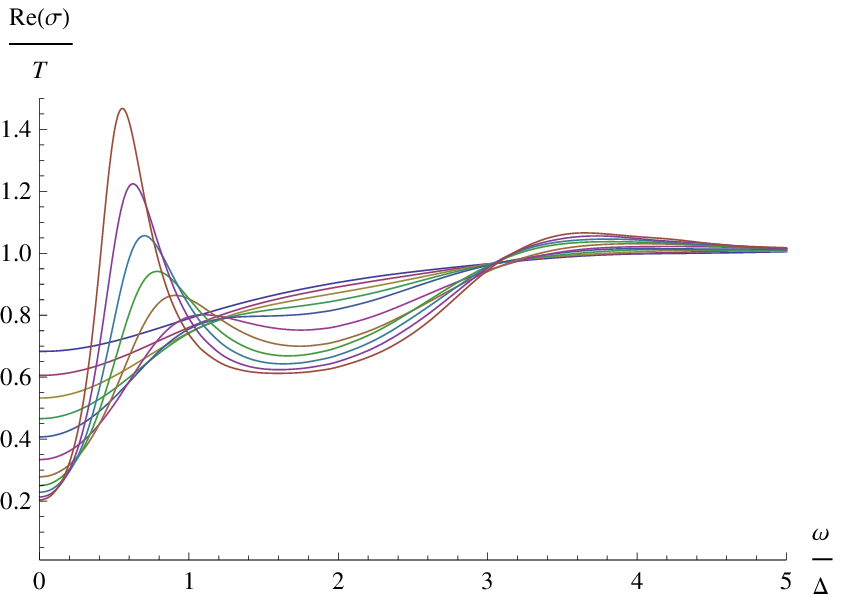}
	\end{minipage}}                
  \subfloat[][]{
	\begin{minipage}[h]{0.5\linewidth}
	\includegraphics[width=0.9 \linewidth]{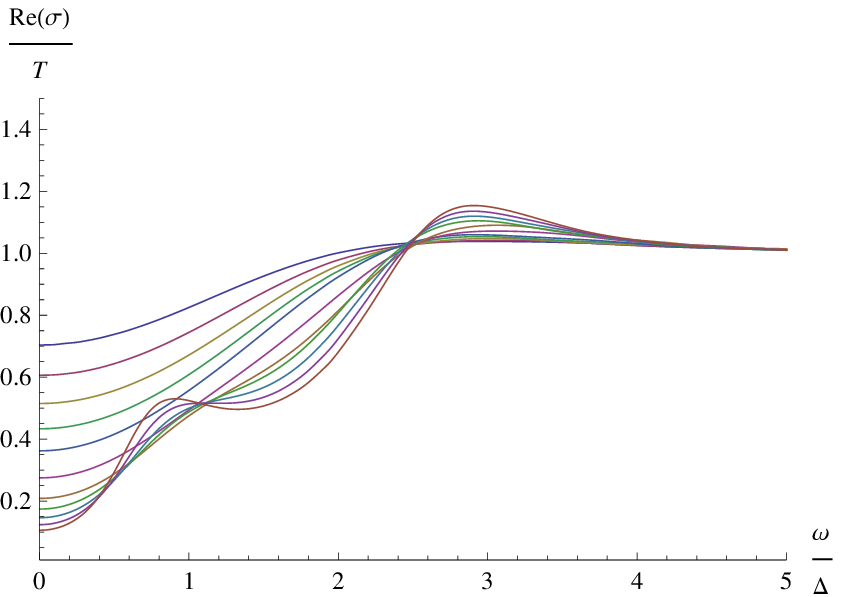}
	\end{minipage}} 

\caption{\label{temps} The real part of the AC conductivity depending on frequency at a) $\lambda=0$ and b) $\lambda= -0.1 \mu$ and various temperatures: $\frac{T}{T_c} = 0.97, 0.96, 0.94, 0.92, 0.9, 0.87, 0.85, 0.81, 0.77, 0.74, 0.71$ (from top to bottom).  For each plot the frequency $\omega$ is normalized with respect to the gap value $\Delta$ at given temperature.}
\end{figure}

To study the dependence of the form of the AC conductivity curve on the gap value, we plot on Fig. \ref{temps} several curves at different temperatures with respect to the frequency, normalized to the corresponding gap value. At $\lambda = 0$ (Fig. \ref{temps}a) we see that the peak position shifts slightly with the temperature, what means that it is affected by the value of the interband current $J$ (Fig. \ref{current}). Another interesting feature is the isotermal point at $\omega \approx 3.1 \Delta$, where the conductivity remains constant at any temperature. Plotting the same set of curves at nonzero external interband potential $\lambda = - 0.1 \mu$ (Fig. \ref{temps}b) we see that this point shifts to $\omega \approx 2.5 \Delta$. Apart of that, the effect of turning on external potential is again in dumping the mid-infrared peak at $\omega \sim \Delta$ and growing the hump at $\omega > 2.5 \Delta$.

Our results are in a close agreement with the calculation of the conductivity of $S^\pm$ state within the Eliashberg theory framework \cite{Golubov}, where the very similar mid-infrared peak was observed. Moreover, in \cite{Golubov} it was discussed that the interband scattering can be controlled by the impurities. This statement is in strong relation with our model in case of nonzero $\lambda$. The effects of impurities can be described in holographic models by introducing external sources for vector operators \cite{impurities}, so we find that introducing nonzero $\lambda$ can be indeed treated as an impurity effect.  

In the end of this section we should stress the importance of the different sign of the electric charge of carriers (electrons and holes) at different conductive bands of our superconductor. Indeed, if the carriers had the same charge sign on both bands, the electromagnetic current would couple to the Abelian part of the gauge field and would not interact with other nonabelian components in the bulk. In such situation we would obtain an equation for electromagnetic fluctuation equivalent to the case of ordinary one-band superconductor \cite{hartnoll1} and would not see any interesting features in the AC conductivity (it would look like the dashed curve on Fig.\ref{conduct}). Hence, the fact that one band is hole-type and another is electron-type is crucial.

\section{Conclusion}

In this work we've constructed the holographic model of a two-band isotropic superconductor. The model includes a charged scalar field in the fundamental representation of $U(2)$ gauge group. The vacuum expectation value of the scalar describes the condensates on the different bands of the superconductor. We found that the pattern of condensation is crucially dependent on the perturbation that resolves the degeneracy of all possible ways of spontaneous breaking of gauge symmetry by the scalar field. We observe the condensation with equal gaps on two bands in the situation, when the effect of perturbation can be expressed as the nonzero off-diagonal chemical potential, which describes the interband interaction. Moreover, we find that the final state, in which the system condenses (which has the maximal value of the critical temperature), depends on the sign of this interband interaction and if the off-diagonal chemical potential $\lambda$ is negative, the $S^\pm$ state is more favorable. This allows us to argue that the next step in development of $S^\pm$ holographic model would be to introduce some additional ingredients (possibly the magnetic subsystem, described by the spin fluctuations, or impurities) in order to generate such an interaction (off-diagonal chemical potential) and resolve the degeneracy without the need of external potentials.  

In our study of the AC electric conductivity we find that the  consequence of the multiband nature of the material is an emergent mid-infrared peak in the real part of the spectra. Although it is suppressed by the same interband potential, which favors the $S^{\pm}$ condensation pattern, and can be melt completely. Moreover, we find some other interesting qualitative features of the AC conductivity spectra, namely the isotermal point and the growing hump right after it. We argue, that the form of spectra we obtain for sufficiently large interband potential $\lambda$ (see Fig. \ref{temps}b), which exhibits the linear rise of the conductivity in mid-infrared region and the subsequent hump, can be related to the real experimental data.

Although the current state of the holographic model do not allow us to make any reliable quantitative predictions, we consider the close qualitative agreement with the results of Eliashberg approach \cite{Golubov} as a strong check of the validity of our model. The further development of the holographic approach would include the introduction of the lattice, impurities and magnetic subsystem, what will eventually provide us with the fully nonperturbative model of superconductivity, which would be presumably of use in high temperature superconductors, where the strong coupling and the absence of small parameters render the applicability of the more conventional approaches questionable.

\acknowledgments
We would like to thank Alexander Gorsky, Andrey Chubukov, Igor Mazin, Aliaksey Charnukha, Alexander Solontsov, Leonid Bork and Stanislav Ogarkov for helpful discussions and comments.

The work of A.K. is partially supported by RFBR grant no.12-02-00284 and PICS- 12-02-91052, the Ministry of Education and Science of the Russian Federation under contract 14.740.11.0347 and the Dynasty Foundation. 

The work of A.V.S. and V.P.K. was supported by the Russian
Ministry of Science and Education under contract No.
 07.514.12.4028 and by the Grant RFBR-11-02-01227-a of
 the Russian Foundation for Basic Research.

\section{Appendix A: equations of motion}

In appendix we consider the full set of the equations of motion of our model (\ref{action}) and check that the ansatz (\ref{ansatz}) satisfies them. First of all introduce the parameterization of the scalar field in the fundamental representation of $U(2)$
\begin{equation}
\phi = \phi_0 e^{i \alpha} \left( \begin{array}{l} \sin(\theta) e^{i \gamma} \\ \cos(\theta) e^{-i \gamma} \end{array} \right)
\end{equation}
and the $U(2)$ gauge field
\begin{equation}
A_\mu = \hat{A}_\mu  \begin{pmatrix} 1 & 0 \\ 0 & 1\end{pmatrix} + A_\mu^1 \begin{pmatrix} 0 & 1 \\ 1 & 0\end{pmatrix} +  A_\mu^2 \begin{pmatrix} 0 & i \\ -i & 0\end{pmatrix}+  A_\mu^3 \begin{pmatrix} 1 & 0 \\ 0 & -1\end{pmatrix}.
\end{equation}
The action of the model (\ref{action}) in this notation takes the form
\begin{align}
S = \int d^3 x dr \ \sqrt{-g} \Bigg[& - \frac{1}{2} \hat{F}_{\mu \nu} \hat{F}^{\mu \nu} - \frac{1}{2} F_{\mu \nu}^i F^{i \ \mu \nu} - 4 e^{ijk} \p_\mu A^i_\nu \ A^{j \ \mu} A^{k \ \nu} \\
& - 2 (\delta_{jl} \delta_{km} - \delta_{jm} \delta_{kl}) A_\mu^j A_\nu^k A^{l \ \mu} A^{m \ \nu} \\
& - (\p_\mu \phi_0)^2 - \phi_0^2 \big(\hat{A}_\mu + \p_\mu \alpha \big)^2 - \phi_0^2 \big(A_\mu^3 + \p_\mu \gamma \big)^2 \\
& +2 \phi_0^2 \cos(2\theta) g^{\mu \nu}\big(\hat{A}_\mu + \p_\mu \alpha \big)\big(A^{3}_{\nu}+ \p_\nu \gamma \big) \\
& - 2 \phi_0^2 \sin(2\theta) g^{\mu \nu} \big(\hat{A}_\mu + \p_\mu \alpha \big) \big(A_\nu^1 \cos(2 \gamma) + A_\nu^2 \sin(2 \gamma) \big) \\
& - 2 \phi_0^2  \big(A_\mu^1 \sin(2 \gamma) - A_\mu^2 \cos(2 \gamma) \big) \p^{\mu} \theta - \phi_0^2 \Big( (A^1_\mu)^2 + (A^2_\mu)^2 \Big) \\
& - \phi_0^2 (\p_\mu \theta)^2  + 2 \phi_0^2  \Bigg],
\end{align}
where $\hat{F} = d \hat{A}$ and $F^i = d A^i$. From this action we derive the equations of motion in an arbitrary gauge:

\begin{align*}
\phi_0: \qquad  \frac{1}{\sqrt{-g}} \p_\mu \sqrt{-g} & g^{\mu \nu} \p_\nu \phi_0 - \phi_0 \Big[ (\hat{A}_\mu + \p_\mu \alpha \big)^2 + \big(A_\mu^3 + \p_\mu \gamma \big)^2 \\
& - 2 \cos(2\theta) g^{\mu \nu}\big(\hat{A}_\mu + \p_\mu \alpha \big)\big(A^{3}_{\nu}+ \p_\nu \gamma \big) \\
& + 2 \sin(2\theta) g^{\mu \nu} \big(\hat{A}_\mu + \p_\mu \alpha \big) \big(A_\nu^1 \cos(2 \gamma) + A_\nu^2 \sin(2 \gamma) \big) \\
& + 2 \big(A_\mu^1 \sin(2 \gamma) - A_\mu^2 \cos(2 \gamma) \big) \p^{\mu} \theta + \Big( (A^1_\mu)^2 + (A^2_\mu)^2 \Big) + (\p_\mu \theta)^2 -2 \Big] = 0, 
\\
\alpha: \qquad  \frac{1}{\sqrt{-g}} \p_\mu \sqrt{-g} & g^{\mu \nu} \Big[\phi_0^2 \big(\hat{A}_\nu + \p_\nu \alpha \big) - \phi_0^2 \cos(2\theta) \big(A^{3}_{\nu}+ \p_\nu \gamma \big) \\ 
& + \phi_0^2 \sin(2\theta) \big(A_\nu^1 \cos(2 \gamma) + A_\nu^2 \sin(2 \gamma) \big)\Big] = 0, 
\\
\gamma: \qquad  \frac{1}{\sqrt{-g}} \p_\mu \sqrt{-g} & g^{\mu \nu} \Big[ \phi_0^2 \big(A^{3}_{\nu}+ \p_\nu \gamma \big) - \phi_0^2 \cos(2\theta) \big(\hat{A}_\nu + \p_\nu \alpha \big) \Big] \\
& + 2 \phi_0^2 \sin(2\theta) g^{\mu \nu} \big(\hat{A}_\mu + \p_\mu \alpha \big) \big( A_\nu^1 \sin(2 \gamma) -  A_\nu^2 \cos(2 \gamma) \big) \\
& - 2 \phi_0^2  g^{\mu \nu} \big(A_\mu^1 \cos(2 \gamma) + A_\mu^2 \sin(2 \gamma) \big) \p_{\nu} \theta = 0, 
\\
\theta: \qquad  \frac{1}{\sqrt{-g}} \p_\mu \sqrt{-g} & g^{\mu \nu} \Big[ \phi_0^2 \p_\nu \theta + \phi_0^2  \big(A_\nu^1 \sin(2 \gamma) - A_\nu^2 \cos(2 \gamma) \big) \Big] \\
& - 2 \phi_0^2 \sin(2\theta) g^{\mu \nu}\big(\hat{A}_\mu + \p_\mu \alpha \big)\big(A^{3}_{\nu}+ \p_\nu \gamma \big) \\
& - 2 \phi_0^2 \cos(2\theta) g^{\mu \nu} \big(\hat{A}_\mu + \p_\mu \alpha \big) \big(A_\nu^1 \cos(2 \gamma) + A_\nu^2 \sin(2 \gamma) \big) = 0,
\\
\hat A_\mu: \qquad \frac{1}{\sqrt{-g}} \p_\nu \sqrt{-g} & g^{\mu \rho} g^{\nu \lambda}  \hat{F}_{\lambda \rho} -  \phi_0^2 g^{\mu \rho} \Big( \hat{A}_\rho + \p_\rho \alpha \Big) \\
& +  \phi_0^2 \cos(2 \theta) g^{\mu \rho} \big(A^{3}_{\rho}+ \p_\rho \gamma \big) \\
& -  \phi_0^2 \sin(2\theta)  g^{\mu \rho} \big(A_\rho^1 \cos(2 \gamma) + A_\rho^2 \sin(2 \gamma) \big) = 0, 
\\
A^1_\mu:    \qquad \frac{1}{\sqrt{-g}} \p_\nu \sqrt{-g} & g^{\mu \rho} g^{\nu \lambda} \Big[ F^1_{\lambda \rho} + 2 \big(A^2_\lambda A^3_\rho - A^3_\lambda A^2_\rho  \big) \Big] - 2 g^{\mu \rho} g^{\nu \lambda} \big(F^3_{\rho \nu} A^2_\lambda - F^2_{\rho \nu} A^3_\lambda \big) \\
& - 4 g^{\mu \rho} g^{\nu \lambda} \Big[ A_\rho^1 (A_\nu^i A_\lambda^i) - A_\lambda^1 (A_\rho^i A_\nu^i)  \Big] \\
& -  \phi_0^2 \sin(2 \theta) \cos(2 \gamma)  g^{\mu \rho} \big(\hat{A}_\rho + \p_\rho \alpha \big) -  \phi_0^2 \sin(2 \gamma)  g^{\mu \rho} \p_\rho \theta -  \phi_0^2 g^{\mu \rho} A^1_\rho =0, 
\\
A^2_\mu:    \qquad \frac{1}{\sqrt{-g}} \p_\nu \sqrt{-g} & g^{\mu \rho} g^{\nu \lambda} \Big[ F^2_{\lambda \rho} + 2 \big(A^3_\lambda A^1_\rho - A^1_\lambda A^3_\rho  \big) \Big] - 2 g^{\mu \rho} g^{\nu \lambda} \big(F^1_{\rho \nu} A^3_\lambda - F^3_{\rho \nu} A^1_\lambda \big) \\
& - 4 g^{\mu \rho} g^{\nu \lambda} \Big[ A_\rho^2 (A_\nu^i A_\lambda^i) - A_\lambda^2 (A_\rho^i A_\nu^i)  \Big] \\
& -  \phi_0^2 \sin(2 \theta) \sin(2 \gamma)  g^{\mu \rho} \big(\hat{A}_\rho + \p_\rho \alpha \big) +  \phi_0^2 \cos(2 \gamma)  g^{\mu \rho} \p_\rho \theta -  \phi_0^2 g^{\mu \rho} A^2_\rho =0, 
\\
A^3_\mu:    \qquad \frac{1}{\sqrt{-g}} \p_\nu \sqrt{-g} & g^{\mu \rho} g^{\nu \lambda} \Big[ F^3_{\lambda \rho} + 2 \big(A^1_\lambda A^2_\rho - A^2_\lambda A^1_\rho  \big) \Big] - 2 g^{\mu \rho} g^{\nu \lambda} \big(F^2_{\rho \nu} A^1_\lambda - F^1_{\rho \nu} A^2_\lambda \big) \\
& - 4 g^{\mu \rho} g^{\nu \lambda} \Big[ A_\rho^3 (A_\nu^i A_\lambda^i) - A_\lambda^3 (A_\rho^i A_\nu^i)  \Big] \\
& + \phi_0^2 \cos (2 \theta) g^{\mu \rho} \big(\hat{A}_\rho + \p_\rho \alpha \big)  -  \phi_0^2 g^{\mu \rho} (A^3_\rho + \p_\rho \gamma) =0. 
\end{align*}
Examining the equation for $A^3_\mu$ we find that in the static ansatz with $A^2_\mu = A^3_\mu = 0$, which is used in the paper, it takes the form 
\begin{equation*} 
\phi_0^2 \cos (2 \theta) g^{\mu \rho} \hat{A}_\rho =0.
\end{equation*}
Given nonzero $\hat{A}_\mu$ this equation fixes two possible values of $\theta$ mentioned in (\ref{theta-values}).

It is interesting to find out the relative phase $\gamma$, which fits our ansatz. For $\theta$ from (\ref{theta-values}) the equation of motion for $\gamma$ leads to 
\begin{equation*} 
2 \phi_0^2 \sin(2\theta) g^{\mu \nu} \hat{A}_\mu \big( A_\nu^1 \sin(2 \gamma) -  A_\nu^2 \cos(2 \gamma) \big)  = 0.
\end{equation*}
One can see that in the case $A^1\neq 0$, which we use in the paper, $\gamma$ is fixed to $\frac{\pi}{2} n, n \in \mathbb{Z}$. This corresponds to the condensate proportional to the vectors $(1,1)^T$ or $(1,-1)^T$, which are the eigenvectors of the Pauli matrix $\sigma^1$. From the other hand, taking $A^2\neq 0$ would lead to $\gamma = \frac{\pi}{4} (1+ 2n), n \in \mathbb{Z}$ and the condensate proportional to $(1,i)^T$ or $(1,-i)^T$, which are the eigenvectors of $\sigma^2$. This demonstrates the fact that the scalar field tends to condense in the state, which is described by the eigenvector of the nonabelian gauge field taken nonzero in a particular ansatz.     

Substituting the ansatz (\ref{ansatz}) to the rest of the equations of motion, one can check that they reduce to
\begin{align*}
\phi_0:& \qquad  \frac{1}{\sqrt{-g}} \p_\mu \sqrt{-g}  g^{\mu \nu} \p_\nu \phi_0 - \phi_0 \Big[ g^{\mu \nu} \big(\hat{A}_\mu \pm A_\mu^1 \big)\big(\hat{A}_\nu \pm A_\nu^1 \big) -2 \Big] = 0, 
\\
\alpha:& \qquad  \frac{1}{\sqrt{-g}} \p_\mu \sqrt{-g}  g^{\mu \nu} \phi_0^2 \big(\hat{A}_\nu    \pm A_\nu^1 \big) = 0, 
\\
\hat A_\mu:& \qquad \frac{1}{\sqrt{-g}} \p_\nu \sqrt{-g}  g^{\mu \rho} g^{\nu \lambda}  \hat{F}_{\lambda \rho} -  \phi_0^2 g^{\mu \rho} \Big( \hat{A}_\rho \pm A_\rho^1 \Big)  = 0, 
\\
A^1_\mu:&    \qquad \frac{1}{\sqrt{-g}} \p_\nu \sqrt{-g}  g^{\mu \rho} g^{\nu \lambda}  F^1_{\lambda \rho}   \mp \phi_0^2 g^{\mu \rho} \Big(  \hat{A}_\rho  \pm A^1_\rho \Big) =0. 
\end{align*}
And the equations for $\gamma, \theta, A^2_\mu, A^3_\mu$ are zero.
The equation of motion for $\alpha$ follows from equation for $\hat{A}$, what is an evidence of the gauge symmetry. We see that our ansatz (\ref{ansatz}) satisfies the general equations of motion if  the functions $\phi_0(r), M(r) = \hat{A}_t(r) , \Lambda(r) = A^1_t(r) $ are solutions to (\ref{EOMS}). 

\section{Appendix B: numerical equation of state}

Solving the system (\ref{EOMS2}) numerically we get a one parameter family of solutions $\{B^i(r), \phi^i(r), r_h^i \}$, where 
\begin{align}
B^i(r) = M^i(r) \pm \Lambda^i(r) & \xrightarrow[r\rar \infty]{}  b_0^i r_h^i  - b_1^i \frac{(r_h^i)^2}{r} , \\
\phi^i(r) & \xrightarrow[r\rar \infty]{}  \hat{\phi}^{i} \frac{(r_h^i)^2}{r^2}.
\end{align}
On the other hand one can easily find the solution of (\ref{EOM3}), which satisfies the boundary conditions (\ref{asympt}) and vanishes at the horizon (as due to the divergence of $g^{tt}$ at $r=r_h$ one needs to impose the condition $A_t(r_h)=0$)
\begin{equation}
M^i(r) \mp \Lambda^i(r) = (\mu^i \mp \lambda^i) \left[ 1 - \frac{r_h^i}{r} \right].  
\end{equation}
These relations allow us to derive the corresponding parameters of the $i$-th solution. In the $S^{++}$ case they are
\begin{align}
T^i &= \frac{3}{4 \pi} \sqrt{\frac{2 \rho_0}{b_1^i + \left(\frac{1-\alpha}{1+\alpha}\right)b_0^i}}, & & \mu^i = b_0^i \frac{1}{1+\alpha} \sqrt{\frac{2 \rho_0}{b_1^i + \left(\frac{1-\alpha}{1+\alpha}\right)b_0^i}},\\
\Delta^i &=  \hat{\phi}^{i} \frac{2 \rho_0}{b_1^i + \left(\frac{1-\alpha}{1+\alpha}\right)b_0^i}, & & 
J^i = \rho_0 \ \frac{b_1^i - \left(\frac{1-\alpha}{1+\alpha}\right)b_0^i}{b_1^i + \left(\frac{1-\alpha}{1+\alpha}\right)b_0^i}
\end{align}
The case $S^{\pm}$ differs by the change of sign in front of $\alpha$ and $J$. The critical parameters of the numerical solution are
\begin{equation}
b_0^{cr} \approx b_1^{cr} \approx 4.06.
\end{equation}
Thus we get the critical temperatures for $S^{++}$ and $S^{\pm}$ states:
\begin{align}
\label{critical_T}
T^{++}_{c} &= \frac{3}{4 \pi} \sqrt{\frac{2 \rho_0}{4.06 + 4.06 \left(\frac{1-\alpha}{1+\alpha}\right)}} \approx 0.118 \sqrt{ \rho_0 (1+\alpha)}, \\
\notag
T^{\pm}_{c} &= \frac{3}{4 \pi} \sqrt{\frac{2 \rho_0}{4.06 + 4.06 \left(\frac{1+\alpha}{1-\alpha}\right)}} \approx 0.118 \sqrt{ \rho_0 (1-\alpha)}
\end{align}
(compare them with the result of \cite{hartnoll1}: $T \approx 0.118 \sqrt{\rho_0}$).

\end{document}